\documentclass[a4paper,11pt]{article}
\pdfoutput=1 

\usepackage{jheppub} 

\usepackage[T1]{fontenc} 
\usepackage{amsmath}
\usepackage{color}
\usepackage{graphicx}
\usepackage{graphics}
\usepackage{subcaption}
\usepackage{graphicx}
\usepackage{epstopdf}		

\numberwithin{equation}{section}

\newcommand{\SB}[1]{ [#1] }					
\newcommand{\AB}[1]{ \langle #1 \rangle }	

\newcommand{\eq}{\begin{equation}}
\newcommand{\eqe}{\end{equation}}
\newcommand{\eqa}{\begin{eqnarray}}
\newcommand{\eqae}{\end{eqnarray}}

\title{\boldmath The supersymmetric spinning polynomial}

\author[1]{Jin-Yu Liu}
\author[1]{Zhe-Ming You}

\affiliation[1]{Department of Physics and Astronomy, National Taiwan University, Taipei 10617, Taiwan}

\abstract{
In this paper, we construct the supersymmetric spinning polynomials. These are orthogonal polynomials that serve as an expansion basis for the residue or discontinuity of four-point scattering amplitudes, respecting four-dimensional super Poincare invariance. The polynomials are constructed by gluing on-shell supersymmetric three-point amplitudes of one massive two massless multiplets, and are identified with algebraic Jacobi-polynomials. Equipped with these we construct the supersymmetric EFThedron, which geometrically defines the allowed region of Wilson coefficients respecting UV unitarity and super Poincare invariance.
}

\begin{document} 
\maketitle
\flushbottom

\section{Introduction}
Scattering amplitudes have been a fertile ground for the study of implications of fundamental principles, manifested as constraints on its analytic properties. It's utility is often based on being expressible in terms of purely on-shell degrees of freedom. In four-dimension massless theory, this is reflected in the use of spinor-helicity formalism (see \cite{1308.1697, 1708.03872} for review).    Recently, Arkani-Hamed, Huang and Huang introduced a new on-shell massive spinor helicity formalism \cite{1709.04891},  that manifest the little group covariance of four-dimensional scattering amplitudes.  Their supersymmetric extension were further developed by \cite{1902.07204, 1810.04694, 1902.07205}. For any four-point massless tree-amplitude, its residue can be expanded on an orthogonal polynomial basis, with the precise polynomial depending on the helicity of the external and the spin of exchanged states. For external scalars these are just the Legendre polynomials. For external helicity states, the polynomials are proportional to Algebraic Jacobi Function (AJF) $\mathcal{J}^{\alpha,\beta}_{S}\left(-x \right)$, here referred to as spinning polynomials \cite{1709.04891}
\eq
\boxed{
P_{s}^{\{h_{i}\}}(x)=g(h_{i},s)\mathcal{J}^{\alpha,\beta}_{s}\left(-x \right)}\,,
\eqe
where
\begin{align}
\nonumber g(h_{i},s)=&(-1)^{s+h_{1}-h_{2}-h_{3}-h_{4}}\frac{\sqrt{\left( s+\alpha\right) !(s-\alpha)!(s+\beta)!(s-\beta)!}}{(2s)!}\\
\mathcal{J}^{\alpha,\beta}_{s}\left(-x \right) =& \sqrt{\frac{\left(s+\alpha\right) !(s-\alpha)!}{(s+\beta)!(s-\beta)!}}\left( \frac{1+x}{2}\right) ^{\frac{\alpha+\beta}{2}}\left( \frac{1-x}{2}\right) ^{\frac{\alpha-\beta}{2}}\times J^{\alpha+\beta,\alpha-\beta}_{s-\alpha}\left( -x\right). 
\end{align}
$x=cos(\theta)$ is scattering angle in center of mass frame thus $-1\leq x\leq1$. $\alpha=h_{1}-h_{2}, \,\beta=h_{3}-h_{4}$ with $h_i$  the helicity of external states. The fact that they are expandable on orthogonal polynomials, is a simple reflection of the uniqueness of the three-point amplitude between one massive and two massless legs in four-dimensions. In this paper, we will show that when the theory enjoys super poincare invariance, the residue of the four-point amplitude is expandable on the super polynomials $\mathcal{P}^{H_{i}}_{\mathcal{N},\mathcal{S}}(x,\eta_{i}^{\mathcal{N}})$, each represent the exchange of a spin-$\mathcal{S}$ massive super multiplet. These are nothing but AJF  multiplying Grassmann delta function:
\begin{equation}
\begin{split}
&\boxed{\mathcal{P}^{\{H_{i}\}}_{\mathcal{N},\mathcal{S}}(x,\eta_{i}^{\mathcal{N}})=m^{\mathcal{N}}\delta^{(2\mathcal{N})}(Q^{\dagger}_{}) g(H_{i},\mathcal{S})\mathcal{J}^{H_{1}-H_{2},H_{3}-H_{4}}_{\mathcal{S}}\left(-x \right)
}
\end{split}\, ,
\end{equation}
Instead of using $h_i$ to label helicity in the spinning polynomials, we use capital $H_i$ label the helicity of the super multiplets and $\alpha=H_{1}-H_{2}, \,\beta=H_{3}-H_{4}$ here.  Note that since both the super and spinning polynomials are expressed on the same basis of AJF, combined with the fact super multiplet decompose onto a sum of components, implies algebraic relation amongst the AJFs. This is precisely given by the recurrence relation of Jacobi polynomial
\begin{equation}
\begin{split}
J^{a,b}_{n}(-x)=&\frac{(n+a+b+1)}{(2n+a+b+1)}J^{a+1,b}_{n}(-x)-\frac{(n+b)}{(2n+a+b+1)}J^{a+1,b}_{n-1}(-x).
\end{split}
\end{equation}
An explicit example is shown in the fixing external states on specific component field. We use Grassmann derivative to fix external states and the recurrence relation give us: 
\eq
\begin{split}
\frac{\partial}{\partial\eta_{2}}\frac{\partial}{\partial\eta_{4}}\mathcal{P}^{\{H_{i}\}}_{\mathcal{N}=1,\mathcal{S}}(x,\eta_{i})=&m^2P^{\{ H_{1}, H_{2}{-}\frac{1}{2},H_{3},H_{4}{-}\frac{1}{2}\} }_{\mathcal{S}{+}\frac{1}{2}}\left( x\right)\\
&{+}\frac{m^{2}(\mathcal{S}{-}H_{3}{+}H_{4})(\mathcal{S}{-}H_{1}{+}H_{2})}{\left( 2\mathcal{S}\right) \left( 2\mathcal{S}{+}1\right)  }P^{ \{H_{1}, H_{2}{-}\frac{1}{2},H_{3},H_{4}{-}\frac{1}{2}\} }_{\mathcal{S}{-}\frac{1}{2}}\left( x\right)\, .
\end{split}
\eqe
Said in another way, the algebraic relation amongst the AJFs encodes the possible extension of the Poincare group!

\begin{figure}[h!]
  \centering
  \includegraphics[width=0.4\linewidth]{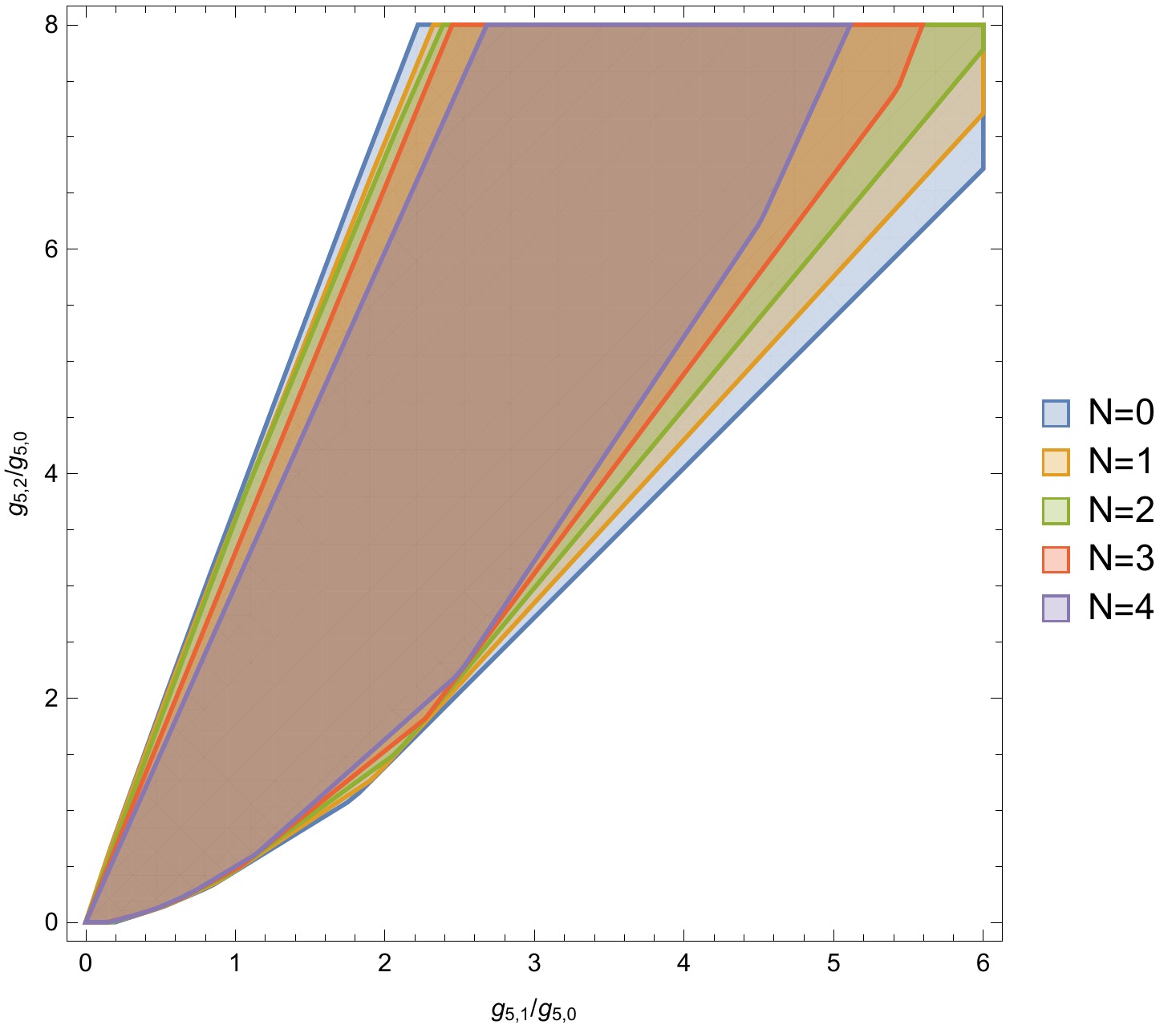}
  \caption{Cyclic-polytope condition on $(g_{5,1}/g_{5,0},g_{5,2}/g_{5,0})$ with supersymmetry $\mathcal{N}$. We don't show $g_{5,3} , g_{5,4}$ here, it will related to $g_{5,1}, g_{5,2}$ by the crossing symmetry $g_{5,i}=g_{5,5-i}$. The bounded region decrease when the superesymmetry $\mathcal{N}$ growth.}
  \label{pin}
\end{figure}

Equipped with the supersymmetric polynomials, we utilize it to constrain the space of EFTs consistent with supersymmetry. As shown in \cite{EFThedron}, unitarity, locality and Poincare invariance allows us to impose bounds on the coefficients of higher dimension operators of the low energy theory. Poincare invaraince is reflected in the fact that the low energy coefficients are bounded by the convex hull of the derivative expansion of the  spinning polynomial. By replacing the latter with supersymmetric polynomial, we obtain new bounds which depends on the number of supersymmetry assumed. We show the bounded region of gluon  amplitude in MHV helicity configuration which can be generally expanded as:
\eq
A^\mathcal{N}(1^+,2^-,3^+,4^-)|_{s,t\ll \Lambda_{UV}}=\SB{1 3}^2\AB{2 4}^2(\frac{a}{st}+\frac{b}{s}{+}\frac{b}{t}+\sum_{k,q\geq 0} g_{k,q}s^{k{-}q}t^q)\,.
\eqe
The convex hull constrains for $g_{k,q}$ can be found in each $k$ order. We show the constrains for each $g_{k,q}$ in supersymmetry between $\mathcal{N}=0\sim4$. The result in $k=5$ is in fig.\ref{pin}. It is directly to see, the higher $\mathcal{N}$ supersymmetry will give us the smaller allowed region.
\\

This paper is organized as follows. In section \ref{sec:Review of spinning polynomials}, we briefly review the spinning polynomials and algebraic Jacobi polynomials. In section \ref{sec:n=1}, we review on-shell $\mathcal{N}=1$ superspace and fix the three-point amplitude with two massless and one massive state consistent with $\mathcal{N}=1$ super Poincare symmetry. We glue two three-point amplitudes and have the residue of $\mathcal{N}=1$ four massless super amplitude. Following the convention of ~\cite{1709.04891}, we transform the result into center of mass frame with scattering angle $cos(\theta)$ and we have $\mathcal{N}=1$ super polynomial. We decompose super polynomial into spinning polynomials by using recurrence relation of AJF.  In section \ref{sec:Extended supersymmetry}, we start from review of $\mathcal{N}=2$ superspace\cite{1902.07204} ,then we derive $\mathcal{N}=2$ super polynomial and the general $\mathcal{N}$ super polynomial. Applying super polynomials on EFT, we show the unitary condition on the low energy coefficients in section \ref{sec:EFT}.
 
\section{Review of spinning polynomials}
\label{sec:Review of spinning polynomials}
Let's start from review spinning polynomials which are the expansion basis for the residues/discontinuities of massless four-point amplitude. The residues of the four-point are nothing but product of two three-point amplitudes. In spacetime dimensional four, two massless one massive amplitudes are uniquely fixed by the helicities $h_{1}$, $h_{2}$ of legs 1 and 2, and the spin-$s$ of the massive leg.
\begin{equation}
\label{}
\begin{split}
A(1^{h_{1}},2^{h_{2}},\textbf{P}^{(I_{1}I_{2}\ldots I_{2s})}) &=m^{1+h_{1}+h_{2}-s}\langle12\rangle^{-h_{1}-h_{2}-s} \prod_{t=1}^{h_{1}-h_{2}+s} \langle2 \textbf{P} ^{(I_{t}}\rangle \prod_{r=h_{1}-h_{2}+s+1}^{2s}\langle1\textbf{P}^{I_{r})}\rangle
\end{split}
\end{equation}
The symmetrized SU(2) indices forms the irreducible representation of spin-$s$ state. The powers of $m$ is to ensure the amplitude has the canonical mass dimension in 4d, i.e. 1. Gluing two vertices, we have the residue of four-massless amplitude with exchanging a spin-$s$ state.
\begin{equation}
\label{gspinpoly}
\begin{split}
Res_{s=m^2}&[A_{4}(h_{i},s)]= A_{L}(1^{h_{1}},2^{h_{2}},\textbf{P}_{(I_{1}I_{2}\ldots I_{2s})})\cdot A_{R}(3^{h_{3}},4^{h_{4}},-\textbf{P}^{(I_{1}I_{2}\cdots I_{2s})})
\\
 &= m^{1+h_{1}+h_{2}-s}
\langle12\rangle^{-h_{1}-h_{2}-s}
\prod_{t=1}^{h_{1}-h_{2}+s}\langle2\textbf{P}_{I_{(t}}\rangle\prod_{r=h_{1}-h_{2}+s+1}^{2s}
\langle 1\textbf{P}_{I_{r)}}\rangle
\\
&\ m^{1+h_{3}+h_{4}-s}
\langle34\rangle^{-h_{3}-h_{4}-s}
\prod_{t=1}^{h_{3}-h_{4}+s}\langle4\vert -\textbf{P}^{I_{(t}}\rangle\prod_{r=h_{3}-h_{4}+s+1}^{2s}
\langle3\vert-\textbf{P}^{I_{r)}}\rangle
\end{split}
\end{equation}
We use convention $\vert{-}\textbf{P}^{I}\rangle=\vert\textbf{P}^{I}\rangle,\,\,\vert{-}\textbf{P}^{I}\rbrack={-}\vert\textbf{P}^{I}\rbrack$ to denote opposite momentum. The tensor contraction of three-point is done by using identity: $\vert \textbf{P}_{I}\rangle_{\alpha}\langle \textbf{P}^{I}\vert^{\beta}=-m_{\textbf{P}}\delta^{\beta}_{\alpha}$. For example, $\left\langle 1\textbf{P}_{I_{r}}\right\rangle $ $\left\langle 3\textbf{P}^{I_{t}}\right\rangle $ can be glued as $\delta^{t}_{r} m_{\textbf{P}}\langle13\rangle$. Then everything we need to do is summing over all contraction in (\ref{gspinpoly}). It is easy to show the combinatorial number in the contraction. Assuming one have contracted spinor $|1\rangle$ and $|3\rangle$ "a" times, then we have the spinor bracket with power 
\eq
\{\langle13\rangle, \langle14\rangle, \langle24\rangle, \langle23\rangle\}=\{a, h_2{-}h_1{+}s{-a}, h_1{-}h_2{+}h_3{-}h_4{+}a, h_{4}{-}h_3{+}s{-}a\}.
\eqe
Then we can write down the combinatorial number for each spinor brackets:
\eqa
\nonumber&\left\langle 13\right\rangle^{a}&:
\left(   
\begin{array}{cr}
  s+h_{2}-h_{1} & \\
  a 
\end{array}\right) \times\left(   
\begin{array}{cr}
  s+h_{4}-h_{3} & \\
  a
\end{array}\right)\times a!\,
\\
\nonumber&\left\langle 14\right\rangle^{(h_{2}-h_{1}+s-a)}&:
\left(\begin{array}{cr}
  h_{3}-h_{4}+s & \\
  h_{2}-h_{1}+s-a
\end{array}\right) \times(h_{2}-h_{1}+s-a)!\,
\\
\nonumber&\left\langle24 \right\rangle^{(h_{1}-h_{2}+h_{3}-h_{4}+a)}&:
\left(   
\begin{array}{cr}
  h_{1}-h_{2}+s & \\
  h_{1}-h_{2}+h_{3}-h_{4}+a
\end{array}\right) \times(h_{1}-h_{2}+h_{3}-h_{4}+a)!\,
\\
\label{eq:permutation}&\left\langle23 \right\rangle^{(h_{4}-h_{3}+s-a)}&:(h_{4}-h_{3}+s-a)!\, .
\eqae
Summing over "a", we have the contraction result in (\ref{gspinpoly}). It can be 
\begin{equation}
\begin{split}
Res_{s=m^2}&[A_{4}(h_{i},s)]=
\\
&\sum_{a}\frac{m^{2-2s+\sum_{i} (h_{i})}}{(2s)!}
\frac{(h_{1}-h_{2}+s)!(h_{2}-h_{1}+s)!(h_{3}-h_{4}+s)!(h_{4}-h_{3}+s)!}{(a)!(s-h_{1}+h_{2}-a)!(s-h_{3}+h_{4}-a)!(h_{1}-h_{2}+h_{3}-h_{4}+a)!}
\\
&\times\left\langle 12\right\rangle^{-h_{1}-h_{2}-s}\left\langle 34\right\rangle^{-h_{3}-h_{4}-s}\langle13\rangle^{a}\langle14\rangle^{s-h_{1}+h_{2}-a}\langle 24\rangle^{h_{1}-h_{2}+h_{3}-h_{4}+a}\langle23\rangle^{h_{4}-h_{3}+s-a} .
\end{split}
\end{equation}
Since the power of spinor brackets are positive integer or zero, the summation is restricted to the region
\eq
a \geq 0,\quad s-h_{1}+h_{2}-a \geq 0, \quad s-h_{3}+h_{4}-a \geq 0, \quad h_{1}-h_{2}+h_{3}-h_{4}+a \geq 0 .
\eqe

It is useful to convert spinor brackets into scattering angle $x=cos\,\theta$ in the center of mass frame. The transformation rule can be found in appendix \ref{sec:spinor-helicity}. As a result, the spinning polynomial is 
\begin{equation}
Res_{s=m^{2}}[A_{4}(h_{i},s)]\equiv P_{s}^{\{h_{i}\}}(x)=g(h_{i},s)\mathcal{J}^{\alpha,\beta}_{s}\left(-x \right)
\end{equation}
where
\begin{align}
\nonumber g(h_{i},s)=&(-1)^{s+h_{1}-h_{2}-h_{3}-h_{4}}\frac{\sqrt{\left( s+\alpha\right) !(s-\alpha)!(s+\beta)!(s-\beta)!}}{(2s)!}\\
\nonumber\mathcal{J}^{\alpha,\beta}_{s}\left(-x \right) =& \sqrt{\frac{\left(s+\alpha\right) !(s-\alpha)!}{(s+\beta)!(s-\beta)!}}\left( \frac{1+x}{2}\right) ^{\frac{\alpha+\beta}{2}}\left( \frac{1-x}{2}\right) ^{\frac{\alpha-\beta}{2}}\times J^{\alpha+\beta,\alpha-\beta}_{s-\alpha}\left( -x\right)\\
&-1 \leq x \leq1 .
\end{align}
Here $\alpha, \beta$ define as $\alpha=h_{1}-h_{2}, \beta=h_{3}-h_{4}$.

Note that the spin-$s$ must satisfy $s\geq \mid h_{1}-h_{2}\mid$ and $\{s, h_{1}-h_{2}\}$ must be all integer or half-integer. $\mathcal{J}^{\alpha,\beta}_{s}$ is Algebraic Jacobi Function(AJF) which satisfy orthogonality condition
\begin{equation}
\begin{array}{lll} 
\int_{-1}^{1}dx\,\, \mathcal{ J}_s^{\alpha,\beta}(x) \mathcal{ J}_{s'}^{\alpha,\beta}(x) = \int_{-1}^{1}dx\,\, (1-x)^{\alpha}(1+x)^{\beta}J^{\alpha,\beta}_{s}(x)J^{\alpha,\beta}_{s'}(x) &=&
\frac{1}{2s+1}\delta_{s\, s'} .
\end{array}
\end{equation}
The advantage of using AJF as opposed to Jacobi-polynomial is the fact that they satisfy the same orthogonality condition as Legendre polynomials.
Said in anotherway, using AJF is equivalent to taking into account the fact that the residue has a universal prefactor that is determined by the external helicities. This relation can be translated to the spinning polynomials
\begin{equation}
\begin{split}
\begin{array}{lll}\label{eq:normalc} 
\int_{-1}^{1}dx\,P_s^{h_{i}}(x)\,\, P_{s'}^{h_{i}}(x) &=&
\frac{\delta_{s\, s'}}{2s+1} \frac{\left( s+\alpha\right) !(s-\alpha)!(s+\beta)!(s-\beta)!}{\left( (2s)!\right)^{2} }.
\end{array}
\end{split}
\end{equation}

\section{$\mathcal{N} = 1$ Supersymmetry}
\label{sec:n=1} 
In this section, we give a brief review of on-shell superspace and superfields. This will allow us to fix the 3-point super amplitude, after which one simply follows the previous example of spinning polynomials: gluing three-point to obtain the super polynomials.
\subsection{Review of $\mathcal{N}=1$ super space} 
We will begin with $\mathcal{N}=1$, the generalization to extended $\mathcal{N}$ are straight forward. Following the convention of \cite{1902.07204} and \cite{0808.1446}, the SUSY generators can be completely written in terms of on-shell variables by introducing grassmann odd variables $\eta$. For example the massive supercharge are
 \begin{equation}
 \begin{split}
q _{\dot{a}}=\tilde{\lambda}_{\dot{a}I}\frac{\partial}{\partial \eta_{I}}\,,\quad   q^{\dagger} _{ a}=\lambda_{a}^{I}\eta_{I} .
\end{split}
 \end{equation}
For massless case, this reduces to
 \begin{equation}
 \begin{split}
q _{\dot{a}}=\tilde{\lambda}_{\dot{a}}\frac{\partial}{\partial \eta}\,,\quad   q^{\dagger} _{ a}=\lambda_{a}\eta .
\end{split}
 \end{equation}
 \\
Supergenerators satisfy super algebra relations, $\{q_{\dot{a}},q_{a}^{\dagger}\}=\textbf{P}_{a\dot{a}}$. For on-shell SUSY representation, we introduce on-shell superfields whose $\eta$ expansion gives the component degrees of freedom. For example, a massive spin-$S$ $\mathcal{N}{=}1$ multiplet can be written as 

\begin{equation}
\label{eq:N1superfield}
 \begin{split}
 \mathcal{S}^{(I_{1}\ldots I_{2S})}&=\phi^{(I_{1}\ldots I_{2S})}+\eta_{J}\Psi^{J(I_{1}\ldots I_{2S})}-\frac{1}{2}\eta_{J}\eta^{J} \tilde{\phi}^{(I_{1}\ldots I_{2S})}\, ,
 \end{split}
 \end{equation}
 where the $(I_{1}\cdots I_{2S})$ denote $\frac{I_{1}I_{2}\cdots I_{2S}+I_{2}I_{1}\cdots I_{2S}+\cdots}{(2S)!}$.
Each order on the expansion gives different spin states with the same mass. The component field  $\Psi^{J(I_{1}\ldots I_{2S})}$ can be separated into irreducible representation by systematically symmetrizing and anti-symmetrizing the free index J with $(I_{1}\cdots I_{2S})$. This leads to $\Psi^{(JI_{1}\ldots I_{2S})}$ and $\Psi_{L}^{\,\, (LI_{2}\ldots I_{2S})}$ representing spin-$S\pm\frac{1}{2}$ states
 
 \begin{equation}
 \Psi^{J(I_{1}\ldots I_{2S})}= \Psi^{(JI_{1}\ldots I_{2S})}+\frac{1}{(2S+1)} \sum_{k=1}^{2S}\epsilon^{JI_{k}} \Psi_{L}^{\,\,(LI_{1}\ldots I_{k-1}I_{k+1}\ldots I_{2S})} \, . 
 \end{equation}
We can derive the component degrees of freedom from (\ref{eq:N1superfield})
\begin{align}
\nonumber\phi^{(I_{1}\ldots I_{2S})}&=\mathcal{S}^{(I_{1}\ldots I_{2S})}\mid_{\eta=0}\\
\nonumber\Psi^{J(I_{1}\ldots I_{2S})}&=\frac{\partial}{\partial \eta_{J}}\mathcal{S}^{(I_{1}\ldots I_{2S})}\mid_{\eta=0}\\
\label{n1de}\tilde{\phi}^{(I_{1}\ldots I_{2S})}&=\frac{1}{2}\frac{\partial}{\partial\eta_{J}}\frac{\partial}{\partial\eta^{J}}\mathcal{S}^{(I_{1}\ldots I_{2S})}\mid_{\eta=0}\, .
\end{align}
For massless multiplet, we instead have
\begin{equation}
 \Phi^{H_{i}}=\phi^{H_{i}}+\eta \psi^{H_{i}-\frac{1}{2}}\, ,
 \end{equation}
 where $\eta$ carries $+\frac{1}{2}$ helicity.\\
 
\noindent\textbf{$\mathcal{N} = 1$ three-point amplitudes}\\
We now consider the three-point super amplitude for the scattering of $H_{1}$, $H_{2}$ massless and a spin-$S$ massive multiplet, $\mathcal{A}(\Phi_1^{H_1},\Phi_1^{H_1},S_{\textbf{p}}^{(I_{1}\ldots I_{2S})})$. The constraint on $\mathcal{A}$ includes it must have $H_1$, $H_2$ helicity weight, carry $(I_1\ldots I_{2s})$ symmetrised indices, and satisfy the supersymmetry Ward identities 
\begin{equation}
Q^{\dagger}_{a}\,\mathcal{A}(\Phi_{1}^{H_{1}},\Phi_{2}^{H_{2}},\mathcal{S}_{\textbf{P}}^{(I_{1}\ldots I_{2S})})=Q_{\dot{a}}\,\mathcal{A}(\Phi_{1}^{H_{1}},\Phi_{2}^{H_{2}},\mathcal{S}_{\textbf{P}}^{(I_{1}\ldots I_{2S})})=0
\end{equation}
where
\begin{equation}
Q^{\dagger}_{ a}=\lambda_{1,a}^{}\eta_{1}+\lambda_{2,a}^{}\eta_{2}+\lambda^{I}_{\textbf{p},a}\eta_{\textbf{P}I}\, .
 \end{equation}
Since $Q_{a}^{\dagger}$ is a multiplicative generator, the amplitude must be proportional to to $\delta^{2}(Q)$
\eq
\mathcal{A}=\delta^{(2)}(Q^{\dagger})f(\lambda,\tilde{\lambda},\eta)\, ,
\eqe
where 
\begin{equation}
\begin{split}
\delta^{(2)}(Q^{\dagger}) = \left( \langle i\vert\eta_{i}+\langle j\vert\eta_{j}+\langle \textbf{P}^{I}\vert\eta_{PI}\right) \left( \vert i\rangle\eta_{i}+\vert j\rangle\eta_{j}+\vert \textbf{P}_{I}\rangle\eta^{PI}\right) 
\\
=\langle1^{}2^{}\rangle\eta_{1^{}}\eta_{2^{}} + \langle1^{}\textbf{P}^{I}\rangle\eta_{1^{}}\eta_{PI} +\langle2^{}\textbf{P}^{I}\rangle\eta_{2^{}}\eta_{PI} +\frac{m}{2}\eta_{PI}\eta_{P}^{I}\, .
\end{split}
\end{equation}
One can then easily check with the momentum conservation
\begin{equation}
Q\delta^{(2)}(Q^{\dagger})=0\, .
\end{equation}
The degree in $\eta$ of $f(\lambda,\tilde{\lambda},\eta)$ for general $\mathcal{N}$ and n-point amplitude is given by
\eq
[ f ]\leq(n-3)\mathcal{N}\, .
\eqe
Since $n=3$, $f(\lambda,\tilde{\lambda},\eta)$ is purely bosonic. Then we have\cite{1902.07204}

\begin{equation}
\label{eq:3pt}
\begin{split}
\mathcal{A}(\Phi_{1}^{H_{1}},\Phi_{2}^{H_{2}},\mathcal{S}_{\textbf{P}}^{(I_{1},\ldots,I_{2S})}) &=\delta^{(2)}(Q^{\dagger})m^{H_{1}+H_{2}-\mathcal{S}}\langle12\rangle^{-H_{1}-H_{2}-\mathcal{S}} \prod_{t=1}^{H_{1}-H_{2}+\mathcal{S}} \langle2\textbf{P}^{(I_{t}}\rangle \prod_{r=H_{1}-H_{2}+\mathcal{S}+1}^{2\mathcal{S}}\langle1\textbf{P}^{I_{r})}\rangle
\end{split}.
\end{equation}
Its' component amplitudes can be extracted by Grassmann derivative. As an example, we show the component amplitude with helicity $H_1{-}\frac{1}{2}$ and $H_2$ on the massless states. There are four states in the massive multiplet and we choose $\Psi^{(I_{1}\ldots I_{2S}J)}, \Theta^{I_{1}\ldots I_{2S}J}$ which both are coming from mixed field $\Psi^{J(I_{1}\ldots I_{2S})}$. Since we are considering mixed field here, we first derive the component amplitude:
\eq
\label{n1cm}
\begin{split}
A(\psi^{H_{1}-\frac{1}{2}},\phi^{H_{2}},\Psi_{\textbf{P}}^{J(I_{1},\ldots,I_{2S})}) &=\frac{\partial }{\partial\eta_{1}}\frac{\partial }{\partial\eta_{PJ}}\mathcal{A}(\Phi^{H_{1}},\Phi^{H_{2}},\mathcal{S}_{\textbf{P}}^{(I_{1},\ldots,I_{2S})})
\\
&=\langle 1\textbf{P}^J\rangle m^{H_{1}+H_{2}-\mathcal{S}}\langle12\rangle^{-H_{1}-H_{2}-\mathcal{S}} \prod_{t=1}^{H_{1}-H_{2}+\mathcal{S}} \langle2\textbf{P}^{(I_{t}}\rangle \prod_{r=H_{1}-H_{2}+\mathcal{S}+1}^{2\mathcal{S}}\langle1\textbf{P}^{I_{r})}\rangle .
\end{split}
\eqe
The mixed field in (\ref{n1cm}) can further decompose onto spin-$\mathcal{S}\pm\frac{1}{2}$ by symmetrizing the SU(2) indices
\begin{equation}
\begin{split}
&A(\psi^{H_{1}-\frac{1}{2}},\phi^{H_{2}},\Psi_{\textbf{P}}^{(I_{1},\ldots,I_{2S}J)})= m^{H_{1}+H_{2}-\mathcal{S}}\langle12\rangle^{-H_{1}-H_{2}-\mathcal{S}} \prod_{t=1}^{H_{1}-H_{2}+\mathcal{S}} \langle2\textbf{P}^{(I_{t}}\rangle \prod_{r=H_{1}-H_{2}+\mathcal{S}+1}^{2\mathcal{S}+1}\langle1\textbf{P}^{I_{r})}\rangle\\
&A(\psi^{H_{1}-\frac{1}{2}},\phi^{H_{2}},\Theta_{\textbf{P}}^{I_{1}\ldots I_{2S}J})= m^{H_{1}+H_{2}-\mathcal{S}}\langle12\rangle^{-H_{1}-H_{2}-\mathcal{S}}
\\
&\times\lbrack\langle 1\textbf{P}^J\rangle\prod_{t=1}^{H_{1}-H_{2}+\mathcal{S}} \langle2\textbf{P}^{(I_{t}}\rangle \prod_{r=H_{1}-H_{2}+\mathcal{S}+1}^{2\mathcal{S}}\langle1\textbf{P}^{I_{r})}\rangle
 -\prod_{t=1}^{H_{1}-H_{2}+\mathcal{S}} \langle2\textbf{P}^{(I_{t}}\rangle \prod_{r=H_{1}-H_{2}+\mathcal{S}+1}^{2\mathcal{S}+1}\langle1\textbf{P}^{I_{r})}\rangle\rbrack .
\end{split}
\end{equation}

\subsection{The N=1 super polynomial}
\label{sec:N=1 spinning polynomial}
With the unique three-point super amplitude, we start to construct the basis for 4-point massless super-amplitude by using $\mathcal{N}=1$ supersum
\begin{equation}
\label{eq:Res}
\begin{split}
Res_{s=m^2}&[\mathcal{A}_{4}(H_{i},\mathcal{S})]= \int \frac{1}{2}\epsilon^{IJ} d\eta_{P_{I}} d\eta_{P_{J}}\mathcal{A}_{L}(\Phi^{H_{1}},\Phi^{H_{2}},\mathcal{S}_{\textbf{P}(I_{1}\cdots I_{2S})})\cdot \mathcal{A}_{R}(\Phi^{H_{3}},\Phi^{H_{4}},\overline{\mathcal{S}}_{-\textbf{P}}^{(I_{1}\cdots I_{2S})})
\\
=\int \frac{1}{2}&\epsilon^{IJ} d\eta_{PI} d\eta_{PJ}
\\
\times\delta^{(2)}&(Q_{L}^{\dagger}) m^{H_{1}+H_{2}-\mathcal{S}}
\langle12\rangle^{-H_{1}-H_{2}-\mathcal{S}}
\prod_{t=1}^{H_{1}-H_{2}+\mathcal{S}}\langle2\textbf{P}_{(I_{t}}\rangle\prod_{r=H_{1}-H_{2}+\mathcal{S}+1}^{2\mathcal{S}}
\langle 1\textbf{P}_{I_{r})}\rangle
\\
\times\delta^{(2)}&(Q_{R}^{\dagger}) m^{H_{3}+H_{4}-\mathcal{S}}
\langle34\rangle^{-H_{3}-H_{4}-\mathcal{S}}
\prod_{t=1}^{H_{3}-H_{4}+\mathcal{S}}\langle4\vert-\textbf{P}^{(I_{t}}\rangle\prod_{r=H_{3}-H_{4}+\mathcal{S}+1}^{2\mathcal{S}}
\langle3\vert-\textbf{P}^{I_{r})}\rangle\,,
\end{split}
\end{equation}
where we identify $\vert{-}\textbf{P}\rangle=\vert\textbf{P}\rangle, \vert {-}\textbf{P}\rbrack={-}\vert \textbf{P}\rbrack, \eta^{I}_{{-}P}={-}\eta^{I}_{P}$. The grassmann delta functions are 
\begin{align}
\nonumber\delta^{(2)}(Q_{L}^{\dagger}) =& \langle1^{}2^{}\rangle\eta_{1^{}}\eta_{2^{}} + \langle1^{}\textbf{P}^{I}\rangle\eta_{1^{}}\eta_{PI} +\langle2^{}\textbf{P}^{I}\rangle\eta_{2^{}}\eta_{PI} +\frac{m}{2}\eta_{PI}\eta_{P}^{I}\\
\delta^{(2)}(Q_{R}^{\dagger}) =& \langle3^{}4^{}\rangle\eta_{3^{}}\eta_{4^{}} + \langle3^{}\vert-\textbf{P}^{I}\rangle\eta_{3^{}}\eta_{PI} +\langle4^{}\vert-\textbf{P}^{I}\rangle\eta_{4^{}}\eta_{PI} +\frac{m}{2}\eta_{PI}\eta_{P}^{I} .
\end{align}
The integration is super sum which only acts on $\delta^{(2)}(Q_{L}^{\dagger})$ and  $\delta^{(2)}(Q_{R}^{\dagger})$. We can reduce Grassmann delta function as
\begin{equation}
\label{eq:eta}
\begin{split}
\int &\frac{1}{2}\epsilon^{AB} d\eta_{PA} d\eta_{PB}\delta^{(2)}(Q_{L}^{\dagger})\delta^{(2)}(Q_{R}^{\dagger})
\\=&m( \langle12\rangle \eta_{1}\eta_{2}+\langle13\rangle \eta_{1}\eta_{3}+\langle14\rangle\eta_{1}\eta_{4}+\langle23\rangle\eta_{2}\eta_{3}+
\langle24\rangle\eta_{2}\eta_{4}+\langle34\rangle\eta_{3}\eta_{4})
\\
=&m\delta^{(2)}(Q^{\dagger}_{4pt})\, .
\end{split}
\end{equation}
As the spinning example (\ref{gspinpoly}), we contract the spinor brackets in (\ref{eq:Res}). The result of the contraction is the same as in (\ref{gspinpoly}) with $h_i\rightarrow H_i$ in (\ref{gspinpoly}). Thus we have:
\begin{equation}
\label{n1sp}
\begin{split}
\boxed{\mathcal{P}^{\{H_{i}\}}_{\mathcal{N}=1,\mathcal{S}}(x,\eta_{i})=m\delta^{(2)}(Q^{\dagger}_{})g_{\left( H_{i},\mathcal{S}\right) }\mathcal{J}^{H_{1}-H_{2},H_{3}-H_{4}}_{\mathcal{S}}\left(-x \right)
}\quad
\end{split},
\end{equation}
where
\begin{equation}
\label{sgp}
g(H_{i},S)=(-1)^{S+H_{1}-H_{2}-H_{3}-H_{4}}\frac{\sqrt{(S+\alpha)!(S-\alpha)!(S+\beta)!(S-\beta)!}}{(2S)!}\, . 
\end{equation}
In this paper, we call this result $\mathcal{P}^{\{H_{i}\}}_{\mathcal{N}=1,\mathcal{S}}(x)$: "$\mathcal{N}=1$ super polynomials". For simplicity, we will use $\alpha,\beta$ denote $H_i$, $\alpha=H_{1}-H_{2},\beta=H_{3}-H_{4}$. The variables $(\mathcal{S},\alpha,\beta)$ must be all integer or all half-integer due to the definition of AJF.\\

\noindent\textbf{Expansion on spinning polynomials}\\
Let us show that $\mathcal{P}^{\{H_{i}\}}_{\mathcal{N}=1,\mathcal{S}}(x,\eta_{i})$ can be recast on to the component spinning polynomials with positive coefficients. To facilitate the comparison between super and spinning polynomial, we first fix the external states of the super polynomial to be of specific component. With the fixed external states, super polynomial can be expanded on spinning polynomial
\begin{equation}
\label{eq:n1 general derivative form}
\frac{\partial}{\partial \eta_{j}}\frac{\partial}{\partial \eta_{k}} \mathcal{P}^{\{H_{i}\}}_{\mathcal{N}=1,\mathcal{S}}(x,\eta_{i})=\sum_{s} m^{2}a^{\{h_{i}\}}_{s} P_{s}^{\{h_{i}\}}(x),\quad j\neq k\, . 
\end{equation}
We can get the coefficient $a_s$ directly by the orthogonality of spinning polynomial
\begin{equation}
\begin{array}{lll} 
\frac{\int_{-1}^{1}dx\,Res_{s=m^2}[ \mathcal{A}_{4}^{(h_{i})}(x)]\,\, P_{s'}^{h_{i}}(x)}{ \frac{( s+\alpha) !(s-\alpha)!(s+\beta)!(s-\beta)!}{( (2s)!)^{2} }} =
\frac{a_{k}}{(2s+1)}
\end{array}\, .
\end{equation}
The same relation can be deduced from the recurrence relation of Jacobi polynomial
\begin{equation}
\label{rcr}
\begin{split}
J^{a,b}_{n}(-x)=\frac{(n+a+b+1)}{(2n+a+b+1)}J^{a+1,b}_{n}(-x)-\frac{(n+b)}{(2n+a+b+1)}J^{a+1,b}_{n-1}(-x)\, .
\end{split}
\end{equation}
With identifying $n=\mathcal{S}{-}\alpha, a=\alpha{+}\beta, b=\alpha{-}\beta$ in (\ref{rcr}), one can also decompose super polynomial. Either way we find
\eq
\label{n1rc}
\begin{split}
\frac{\partial}{\partial\eta_{2}}\frac{\partial}{\partial\eta_{4}}\mathcal{P}^{\{H_{i}\}}_{\mathcal{N}=1,\mathcal{S}}(x,\eta_{i})=&m^2P^{\{ H_{1}, H_{2}{-}\frac{1}{2},H_{3},H_{4}{-}\frac{1}{2}\} }_{\mathcal{S}{+}\frac{1}{2}}\left( x\right)\\
&{+}\frac{m^{2}(\mathcal{S}{-}H_{3}{+}H_{4})(\mathcal{S}{-}H_{1}{+}H_{2})}{\left( 2\mathcal{S}\right) \left( 2\mathcal{S}{+}1\right)  }P^{ \{H_{1}, H_{2}{-}\frac{1}{2},H_{3},H_{4}{-}\frac{1}{2}\} }_{\mathcal{S}{-}\frac{1}{2}}\left( x\right)\, .
\end{split}
\eqe
We have similar relations for other external states which we list in Tab \ref{fig:n1}. 
The first term in the expansion of (\ref{n1rc}) is related to spinning polynomial which have internal state $\Psi^{(I_{1},\ldots,I_{2S}J)}$. Indeed since
\begin{equation}
\begin{split}
 &A_{3}^{\phi^{H_{1}},\psi^{H_{2}-\frac{1}{2}},\Psi^{(I_{1}\ldots I_{2S}J)}}\cdot A_{3}^{\phi^{H_{3}},\psi^{H_{4}-\frac{1}{2}},\tilde{\Psi}_{(I_{1}\ldots I_{2S}J)}}=m^2P^{\{ H_{1}, H_{2}{-}\frac{1}{2},H_{3},H_{4}{-}\frac{1}{2}\} }_{\mathcal{S}{+}\frac{1}{2}}\left( x\right) 
\end{split}
\end{equation}
and
\begin{equation}
\begin{split}
A_{3}^{\phi^{H_{1}},\psi^{H_{2}-\frac{1}{2}},\Theta_{\textbf{P}}^{I_{1}\ldots I_{2S}J}}\cdot A_{3}^{\phi^{H_{3}},\psi^{H_{4}-\frac{1}{2}},\Theta_{\textbf{P}I_{1}\ldots I_{2S}J}}=\frac{m^{2}(\mathcal{S}{-}H_{3}{+}H_{4})(\mathcal{S}{-}H_{1}{+}H_{2})}{\left( 2\mathcal{S}\right) \left( 2\mathcal{S}{+}1\right)  }P^{ \{H_{1}, H_{2}{-}\frac{1}{2},H_{3},H_{4}{-}\frac{1}{2}\} }_{\mathcal{S}{-}\frac{1}{2}}\left( x\right)\, ,
\end{split}
\end{equation}
eq(\ref{n1rc}) is equivalent to
\begin{equation}
\label{eq:14}
\begin{split}
\frac{\partial}{\partial\eta_{2}}\frac{\partial}{\partial\eta_{4}}\mathcal{P}^{\{H_{i}\}}_{\mathcal{N}=1,\mathcal{S}}(x,\eta_{i})
=&A_{3}^{\phi^{H_{1}},\psi^{H_{2}-\frac{1}{2}},\Psi^{(I_{1}\ldots I_{2S}J)}}\cdot A_{3}^{\phi^{H_{3}},\psi^{H_{4}-\frac{1}{2}},\tilde{\Psi}_{(I_{1}\ldots I_{2S}J)}}\\
&+A_{3}^{\phi^{H_{1}},\psi^{H_{2}-\frac{1}{2}},\Theta_{\textbf{P}}^{I_{1}\ldots I_{2S}J}}\cdot A_{3}^{\phi^{H_{3}},\psi^{H_{4}-\frac{1}{2}},\Theta_{\textbf{P}I_{1}\ldots I_{2S}J}}\, .
\end{split}
\end{equation}
Clearly, super polynomial is the super sum of spinning polynomials. Note that the recurrence relation tell us that the super polynomial with $\{H_i, \mathcal{S}\}$ is expanded with spinning polynomials $\{h_i, s\}$ in the region $\left| s-\mathcal{S}_{}\right|\leq\frac{1}{2} \wedge \left| h_{i}-H_{i}\right|\leq\frac{1}{2}$.

\begin{table}[]
\begin{center}
\begin{tabular}{|l|l|l|l|l|l|l|}
\hline
\multicolumn{4}{|l|}{External states helicity} &$a^{\{h_{i}\}}_{s-\frac{1}{2}}$  & $a^{\{h_{i}\}}_{s}$ & $a^{\{h_{i}\}}_{s+\frac{1}{2}}$ \\ \hline
     $H_1{-}\frac{1}{2}$& $H_2{-}\frac{1}{2}$    &   $H_3$ &   $H_4$  & 0 &1  &0  \\ \hline
     $H_1{-}\frac{1}{2}$&  $H_2$   &   $H_3{-}\frac{1}{2}$ &  $H_4$  & $\frac{(\mathcal{S}+H_{3}-H_{4})(\mathcal{S}+H_{1}-H_{2})}{(2\mathcal{S})(2\mathcal{S}+1)}$ & 0 & 1 \\ \hline
\end{tabular}
\caption{\label{fig:n1} The expansion coefficient of super polynomial $\mathcal{P}^{\{H_{i}\}}_{\mathcal{N}=1,\mathcal{S}}$. The difference external states are labeled by helicity $H_i$ and $a^{h_{i}}_{s}$ are the coefficient.}
\end{center}
\end{table}

\section{Extended supersymmetry }
\label{sec:Extended supersymmetry}
Previously we derive $\mathcal{N}=1$ super polynomial which is product of Grassmann delta function and AJF. It is naturally to guess the extended super polynomials are still the product of Grassmann and AJF but Grassmann function should be generalized to $\delta^{(2\mathcal{N})}(Q^{\dagger}_{})$. In this section, we will explicitly show this generalization in $\mathcal{N}=2$ SUSY and list more generalized results in the end. 

\subsection{Review of ${N}=2$ super space}
In $\mathcal{N} = 2$ SUSY, we simply consider superalgebra without central charge. The anti-commutation relation of super charges are
\eq
\left\lbrace q _{\dot{a}A},q^{\dagger B} _{ a}\right\rbrace =\delta^{B}_{A} \textbf{P}_{a\dot{a}}\, ,
\eqe
where
\begin{equation}
\begin{split}
q _{\dot{a}A}=\tilde{\lambda}_{\dot{a}I}\frac{\partial}{\partial \eta^{A}_{I}}\,,\quad   q^{\dagger A} _{ a}=\lambda_{a}^{I}\eta^{A}_{I}\, .
\end{split}
\end{equation}
We use indices $A,B$ denote SU(2) R-symmetry group and $I,J$ denote little group.
In the massless limit, it can reduce to  
\begin{equation}
q _{\dot{a}A}=\tilde{\lambda}_{\dot{a}}\frac{\partial}{\partial \eta^{A}}\,,\quad   q^{\dagger A} _{ a}=\lambda_{a}\eta^{A}\, .
\end{equation}
\\
The $\eta$ expansion of on-shell super field in $\mathcal{N}=2$ is
\begin{equation}
\begin{split}
\mathcal{S}^{\left( I_{1}\ldots I_{2S}\right) }=&\phi^{\left( I_{1}\ldots I_{2S}\right) }+\eta^{A}_{I}\Psi^{ I\left( I_{1}\ldots I_{2S}\right)}_{A}-\frac{1}{2} \eta^A_I \eta^B_J (\epsilon^{IJ} \phi^{\left( I_{1}\ldots I_{2S}\right)}_{(AB)} + \epsilon_{AB} W^{(IJ)\left( I_{1}\ldots I_{2S}\right)})
\\
&+\frac{1}{3}\eta_I^B\eta_{JB}\eta^{JA}\tilde{\Psi}_A^{ I\left(I_{1}\ldots I_{2S}\right)}+\eta^{1}_{1}\eta^{2}_{1}\eta^{1}_{2}\eta^{2}_{2}\tilde{\phi}^{\left( I_{1}\ldots I_{2S}\right)} .
\end{split}
\end{equation}
Under massless limit, the super field can be simplified as
\begin{equation}
\Phi^{H_{i}}= \phi^{H_{i}} +\eta^{A}\, \psi_{A}^{H_{i}-\frac{1}{2}}+\frac{1}{2}\eta^{A}\eta_{A}\, w^{H_{i}-1}.
\end{equation}
We can extrapolate the component field by Grassmann derivative
\begin{align}
\nonumber\phi^{\left( I_{1}\ldots I_{2S}\right)}_{(AB)}&=\frac{1}{2}\frac{\partial}{\partial \eta^{JA}} \frac{\partial}{\partial \eta^{B}_{ J}}\mathcal{S}^{(I_{1}\ldots I_{2S})} \mid_{\eta=0}\\
W^{(IJ)\left( I_{1}\ldots I_{2S}\right)}&=\frac{1}{2}\frac{\partial}{\partial \eta^{A}_{I}} \frac{\partial}{\partial \eta_{AJ}}\mathcal{S}^{(I_{1}\ldots I_{2S})} \mid_{\eta=0}\, ,
\end{align}
where derivative of $\phi^{\left( I_{1}\ldots I_{2S}\right) }$, $\Psi^{ I\left( I_{1}\ldots I_{2S}\right)}_{A}$ were list in $\mathcal{N}=1$.
As $\mathcal{N}=1$ SUSY, component fields under $\eta$ expansion are the mixed states. So we start to decompose component fields into irreducible representation. $\phi^{\left( I_{1}\ldots I_{2S}\right) }$ and $\phi^{\left( I_{1}\ldots I_{2S}\right)}_{(AB)}$ are already irreducible and the decomposition of $\Psi^{ I\left( I_{1}\ldots I_{2S}\right)}_{A}$ can be found in $\mathcal{N}=1$. So we only list decomposition of $W^{(IJ)\left( I_{1}\ldots I_{2S}\right)}$ here. Symmetrizing and anti-symmetrizing the little group indices,  
\begin{equation}
\begin{split}
W^{(IJ)\left( I_{1}\ldots I_{2S}\right)}&=W^{(IJ I_{1}\ldots I_{2S}) }
\\
&+\frac{(2S+1)}{2(2S)(2S+2)}\sum_{k=1}^{2S}\Bigg\lbrack\epsilon^{JI_{k}}(W^{(LI I_{1}\ldots I_{k-1}I_{k+1}\ldots I_{2S})}_{L}
+W^{(L\vert \,\,\,\vert I I_{1}\ldots I_{k-1}I_{k+1}\ldots I_{2S})}_{\,\,\,\,\,\,\,\,\,L}) +I\leftrightarrow J \Bigg\rbrack
\\
&+\frac{1}{2(2S)(2S+1)}\sum_{k\neq j}^{2S}\left[( \epsilon^{II_{k}}\epsilon^{JI_{j}}W^{\quad(LKI_{1}\ldots I_{k-1}I_{k+1}\ldots I_{j-1}I_{j+1}\ldots I_{2S})}_{LK})+I\leftrightarrow J \right] .
\end{split}
\end{equation}
It is directly to see $W^{(IJ)\left( I_{1}\ldots I_{2S}\right)}$ field contain spin-$S\pm1$ and spin-$S$ states. \\

\noindent\textbf{$\mathcal{N}=2$ three-point amplitudes}\\
Following the same procedure in $\mathcal{N}=1$, the three-point amplitude is uniquely fixed in $\mathcal{N}=2$. We list the result with helicity $H_{1}, H_{2}$ and spin-$S$:
\begin{equation}
\begin{split}
&\mathcal{A}(\Phi^{H_{1}},\Phi^{H_{2}},\mathcal{S}_{\textbf{P}}^{\left( I,\ldots,I_{2S}\right) }) 
\\=&\delta^{(4)}(Q^{\dagger})m^{H_{1}+H_{2}-\mathcal{S}}\langle12\rangle^{-H_{1}-H_{2}-\mathcal{S}} \prod_{t=1}^{H_{1}-H_{2}+\mathcal{S}} \langle1\textbf{P}^{( I_{t}}\rangle \prod_{r=H_{1}-H_{2}+\mathcal{S}+1}^{2\mathcal{S}}\langle2\textbf{P}^{I_{r}) }\rangle ,
\end{split}
\end{equation}
where
\begin{equation}
\begin{split}
\delta^{(4)}(Q^{\dagger}) =  \prod^{\mathcal{N}=2}_{A=1}\left( \langle1^{}2^{}\rangle\eta^{A}_{1^{}}\eta^{A}_{2^{}} + \langle1^{}\textbf{P}^{I}\rangle\eta^{A}_{1^{}}\eta^{A}_{PI} +\langle2^{}\textbf{P}^{I}\rangle\eta^{A}_{2^{}}\eta^{A}_{PI} +\frac{m}{2}\eta^{A}_{PI}\eta_{P}^{AI}\right) . 
\end{split}
\end{equation}

\subsection{The $\mathcal{N}=2$ super polynomial}
\label{sec:n2 spinning polynomials}
Having uniquely three-point result in $\mathcal{N}=2$, we start to glue them up. We use super sum to glue them just as what we have done in Sec.\ref{sec:N=1 spinning polynomial}. The difference in $\mathcal{N}=2$ is the super sum should also sum over all the R-symmetry indices
\begin{equation}
\label{eq:Res2}
\begin{split}
Res_{s=m^2}&[\mathcal{A}^{\mathcal{N}=2}(H_{i},\mathcal{S})]= \int \prod_{A=1}^{2}\left( \frac{1}{2} d\eta^{A}_{PI} d\eta_{P}^{AI}\right) \mathcal{A}_{L}(\Phi^{H_{1}},\Phi^{H_{2}},\mathcal{S}_{\textbf{P}(I_{1}\cdots I_{2S})})\cdot \mathcal{A}_{R}(\Phi^{H_{3}},\Phi^{H_{4}},\overline{\mathcal{S}}_{-\textbf{P}}^{(I_{1}\cdots I_{2S})})
\\
=&\int \frac{1}{2^{2}} d\eta^{A}_{PI} d\eta_{P}^{AI}\\
\times\delta^{(4)}&(Q_{L}^{\dagger}) m^{H_{1}+H_{2}-\mathcal{S}}
\langle12\rangle^{-H_{1}-H_{2}-\mathcal{S}}
\prod_{t=1}^{H_{1}-H_{2}+\mathcal{S}}\langle2\textbf{P}_{(I_{t}}\rangle\prod_{r=H_{1}-H_{2}+\mathcal{S}+1}^{2\mathcal{S}}
\langle 1\textbf{P}_{I_{r})}\rangle
\\
\times\delta^{(4)}&(Q_{R}^{\dagger}) m^{H_{3}+H_{4}-\mathcal{S}}
\langle34\rangle^{-H_{3}-H_{4}-\mathcal{S}}
\prod_{t=1}^{H_{3}-H_{4}+\mathcal{S}}\langle4\vert-\textbf{P}^{(I_{t}}\rangle\prod_{r=H_{3}-H_{4}+\mathcal{S}+1}^{2\mathcal{S}}
\langle3\vert-\textbf{P}^{I_{r})}\rangle .
\end{split}
\end{equation}
The $\eta$ integration only hit on Grassmann delta function and the integral result is
\begin{equation}
\label{eq:supersumn2}
\begin{split}
&\int \frac{1}{2^{2}}\prod_{A=1}^{2} \left( d\eta^{A}_{PI} d\eta_{P}^{AI} \right) \delta^{(4)}(Q_{L}^{\dagger})\delta^{(4)}(Q_{R}^{\dagger})
\\
&=m^{2}\prod_{A=1}^{2}( \langle12\rangle \eta^{A}_{1}\eta^{A}_{2}+\langle13\rangle \eta^{A}_{1}\eta^{A}_{3}+\langle14\rangle\eta^{A}_{1}\eta^{A}_{4}+\langle23\rangle\eta^{A}_{2}\eta^{A}_{3}+
\langle24\rangle\eta^{A}_{2}\eta^{A}_{4}+\langle34\rangle\eta^{A}_{3}\eta^{A}_{4})
\\
&=m^{2}\delta^{(4)}(Q^{\dagger}) .
\end{split}
\end{equation}
As the same procedure in spinning polynomial (\ref{gspinpoly}), we contract spinor brackets and have
\begin{equation}
\label{eq:4pt2}
\begin{split}
&\boxed{\mathcal{P}^{\{H_{i}\}}_{\mathcal{N}=2,\mathcal{S}}(x,\eta_{i}^{\mathcal{N}})=m^{2}\delta^{(4)}(Q^{\dagger}) g(H_{i},\mathcal{S})\mathcal{J}^{H_{1}-H_{2},H_{3}-H_{4}}_{\mathcal{S}}\left(-x \right)
}
\end{split}
\end{equation}
where $g(H_{i},\mathcal{S})$ is defined in (\ref{sgp}). As we expected, $\mathcal{N}=2$ super polynomial is still product of Grassmann delta function with AJF.\\

\noindent\textbf{Expansion on spinning polynomial}\\
Using the recurrence relation help us to decompose super polynomial into spinning polynomials in $\mathcal{N}=1$. Here we will use the recurrence relation again and show the decomposition in $\mathcal{N}=2$. The only difference between $\mathcal{N}=1$ and $\mathcal{N}=2$ is we need to repeat the recurrence relation twice in $\mathcal{N}=2$ and $P^{\{H_{i}\}}_{\mathcal{N}=2,\mathcal{S}}$ will be decomposed into four elements. The recurrence relation for Jacobi polynomial is:
\begin{equation}
\label{eq:1324recurrence relationJ}
\begin{split}
J^{a,b}_{n}(x)=& \frac{(n+a+b+1)}{(2n+a+b+1)}J^{a,b+1}_{n}(x)+\frac{(n+a)}{(2n+a+b+1)}J^{a,b+1}_{n-1}(x)\\
=&\frac{(n+a+b+1)}{(2n+a+b+1)}\frac{1}{(n+\frac{a}{2}+\frac{b}{2}+1)(1+x)}[(n+1)J^{a,b}_{n+1}(x)+(n+b+1)J^{a,b}_{n}(x)]\\
&+\frac{(n+a)}{(2n+a+b+1)}\frac{1}{(n+\frac{a}{2}+\frac{b}{2})(1+x)}[(n)J^{a,b}_{n}(x)+(n+b)J^{a,b}_{n-1}(x)]\, ,
\end{split}
\end{equation}
Applying (\ref{eq:1324recurrence relationJ}) into AJF, we can simply decompose super polynomial.  As an example, we consider specific component amplitude here
\eq
\frac{\partial}{\partial\eta^{A}_{1}}\frac{\partial}{\partial\eta^{B}_{2}}\frac{\partial}{\partial\eta^{A}_{3}}\frac{\partial}{\partial\eta^{B}_{4}}\mathcal{P}_{\mathcal{N}=2,\mathcal{S}}^{\{H_{i}\}}(x,\eta^{C}_{i})\, .
\eqe
Identifying $n=\mathcal{S}-\alpha, a=\alpha{+}\beta, b=\alpha{-}\beta$ in recurrence relation, we decompose the component amplitude into
\begin{equation}
\label{N2ex}
\begin{split}
&\frac{\partial}{\partial\eta^{A}_{1}}\frac{\partial}{\partial\eta^{B}_{2}}\frac{\partial}{\partial\eta^{A}_{3}}\frac{\partial}{\partial\eta^{B}_{4}}\mathcal{P}_{\mathcal{N}=2,\mathcal{S}}^{\{H_{i}\}}(x,\eta^{C}_{i})=
\\
&m^{4}\{ P^{ \{H_{i}-\frac{1}{2}\} }_{\mathcal{S}+1}\left( x\right)+\Big(\frac{\left(\mathcal{S}+\beta \right) \left(\mathcal{S}-\alpha \right)}{\left( 2\mathcal{S}\right) \left( 2\mathcal{S}+1\right) }+\frac{\left(\mathcal{S}-\beta+1 \right) \left(\mathcal{S}+\alpha+1 \right)}{\left( 2\mathcal{S}+1\right) \left( 2\mathcal{S}+2\right) }\Big)P^{ \{H_{i}-\frac{1}{2}\} }_{\mathcal{S}}\left( x\right)
\\
&+\frac{\left(\mathcal{S}+\beta \right)\left(\mathcal{S}-\beta \right) \left(\mathcal{S}+\alpha \right)\left(\mathcal{S}-\alpha \right)P^{ \{H_{i}-\frac{1}{2}\} }_{\mathcal{S}-1}\left( x\right)}{\left( 2\mathcal{S}-1\right)\left( 2\mathcal{S}\right)^2 \left( 2\mathcal{S}+1\right) }\}\, .
\end{split}
\end{equation}
Similar results for other external states are in Tab.\ref{fig:n2} where
\begin{equation}
\label{N2cc}
\begin{split}
 \mathfrak{a}&= \frac{(\mathcal{S}+\alpha)(\mathcal{S}+\alpha-1)(\mathcal{S}-\beta)(\mathcal{S}-\beta-1)}{(2\mathcal{S}-1)(2\mathcal{S})^{2}(2\mathcal{S}+1)}
 \\
 \mathfrak{b}&=\frac{(\mathcal{S}+\alpha)(\mathcal{S}-\beta)}{(2\mathcal{S})(2\mathcal{S}+1)}+\frac{(\mathcal{S}+\alpha)(\mathcal{S}-\beta)}{(2\mathcal{S}+1)(2\mathcal{S}+2)}
 \\
 \mathfrak{c}&=\frac{\left(\mathcal{S}+\beta \right)\left(\mathcal{S}-\beta \right) \left(\mathcal{S}+\alpha \right)!}{\left(\mathcal{S}+\alpha-2 \right)!\left( 2\mathcal{S}-1\right)\left( 2\mathcal{S}\right)^2 \left( 2\mathcal{S}+1\right) }
 \\
 \mathfrak{d}&=\frac{\left(\mathcal{S}+\beta \right) \left(\mathcal{S}+\alpha \right)}{\left( 2\mathcal{S}\right) \left( 2\mathcal{S}+1\right) }
+\frac{\left(\mathcal{S}-\beta+1 \right) \left(\mathcal{S}+\alpha \right)}{\left( 2\mathcal{S}+1\right) \left( 2\mathcal{S}+2\right) }
 \\
 \mathfrak{e}&=\frac{(\mathcal{S}+\alpha)(\mathcal{S}-\beta)}{(2\mathcal{S})(2\mathcal{S}+1)}
 \\
 \mathfrak{f}&=\frac{(\mathcal{S}+\alpha)(\mathcal{S}-\alpha)(\mathcal{S}+\beta)(\mathcal{S}-\beta)}{(2\mathcal{S}-1)(2\mathcal{S})^{2}(2\mathcal{S}+1)}
 \\
 \mathfrak{g}&=\frac{(\mathcal{S}-\alpha)(\mathcal{S}+\beta)}{(2\mathcal{S})(2\mathcal{S}+1)}+\frac{(\mathcal{S}+\alpha+1)(\mathcal{S}-\beta+1)}{(2\mathcal{S}+1)(2\mathcal{S}+2)} .
 \end{split}
\end{equation}

\begin{table}[]
\begin{tabular}{|l|l|l|l|l|l|l|l|l|}
\hline
\multicolumn{4}{|c|}{External states helicity} & $a^{\{h_{i}\}}_{s-1}$ &$a^{\{h_{i}\}}_{s-\frac{1}{2}}$  &$a^{\{h_{i}\}}_{s}$  & $a^{\{h_{i}\}}_{s+\frac{1}{2}}$ & $a^{\{h_{i}\}}_{s+1}$ \\ \hline
     $(H_1{-}1)_{AB}$&  $(H_2{-}1)_{AB}$   & $H_3$    &  $H_4$   & 0 & 0 & $\epsilon^{BA}\epsilon^{BA}$ & 0 &0  \\ \hline
     $(H_1{-}1)_{AB}$&  $H_2$   & $H_3$    &  $(H_4{-}1)_{AB}$   & $\mathfrak{a}\epsilon^{BA}\epsilon^{BA}$ & 0 &$\mathfrak{b}\epsilon^{BA}\epsilon^{BA}$  & 0 & $\epsilon^{BA}\epsilon^{BA}$ \\ \hline
    $(H_1{-}1)_{AB}$ &   $H_2$  &  $(H_3{-}\frac{1}{2})_{A}$   & $(H_4{-}\frac{1}{2})_{B}$    & $\mathfrak{c}\epsilon^{BA}$ & 0 & $\mathfrak{d}\epsilon^{BA}$ & 0 &$\epsilon^{BA}$  \\ \hline
    $(H_1{-}1)_{AB}$ &  $(H_2{-}\frac{1}{2})_{A}$   & $H_3$    &  $(H_4{-}\frac{1}{2})_{B}$   & 0 &$\mathfrak{e}\epsilon^{BA} $ &0  &$\epsilon^{BA}$  & 0 \\ \hline
    $(H_1{-}\frac{1}{2})_{A}$ &  $(H_2{-}\frac{1}{2})_{A}$   &  $(H_3{-}\frac{1}{2})_{B}$&  $(H_4{-}\frac{1}{2})_{B}$ & 0 &0  &1  &0  &0  \\ \hline
    $(H_1{-}\frac{1}{2})_{A}$ & $(H_2{-}\frac{1}{2})_{B}$    &  $(H_3{-}\frac{1}{2})_{A}$& $(H_4{-}\frac{1}{2})_{B}$     & $\mathfrak{f}$ & 0 & $\mathfrak{g}$ &0  &1  \\ \hline
\end{tabular}
\caption{\label{fig:n2} Super polynomial $\mathcal{P}^{\{H_{i}\}}_{\mathcal{N}=2,\mathcal{S}}(x,\eta_{i}^{C})$ decomposed into spinning polynomials by recurrence relation. We use helicity $H_i$ to label each component states and the indices A,B are R indices in $\mathcal{N}=2$. Number $\mathfrak{a}, \mathfrak{b}, \mathfrak{c}, ......$ in the coefficient $a^{\{h_{i}\}}_{s}$ list in eq(\ref{N2cc}). }
\end{table}
The decomposition of (\ref{N2ex}) are precisely correspond to sum over spinning polynomials with internal states: $\phi_{\textbf{P}(CD)}^{(I_{1},\ldots,I_{2S})}$, $\mathcal{W}_{S}$ and $\mathcal{W}_{S\pm1}$ where
\begin{align}
\nonumber\mathcal{W}_{S+1}=&W_{\textbf{P}}^{(IJI_{1},\ldots,I_{2S})}\\
\nonumber\mathcal{W}_{S}=&\frac{(2S+1)}{2(2S)(2S+2)}\sum_{k=1}^{2S}\left[\epsilon^{JI_{k}}(W^{\,(LI I_{1}\ldots I_{k-1}I_{k+1}\ldots I_{2S})}_{L}+W^{(L\vert \,\,\,\vert I I_{1}\ldots I_{k-1}I_{k+1}\ldots I_{2S})}_{\,\,\,\,\,\,\,\,\,L}) +I\leftrightarrow J\right]\\
\nonumber\mathcal{W}_{S-1}=&\frac{1}{2(2S)(2S+1)}\sum_{k \neq j}^{2S}\left[( \epsilon^{II_{k}}\epsilon^{JI_{j}}W^{\quad(LKI_{1}\ldots I_{k-1}I_{k+1}\ldots I_{j-1}I_{j+1}\ldots I_{2S})}_{LK})+I\leftrightarrow J\right] .
\end{align}
Eq(\ref{N2ex}) is equivalent to
\begin{equation}
\label{eq:N=2,13A24B split to RtripletRsinglet}
\begin{split}
&\frac{\partial}{\partial\eta^{A}_{1}}\frac{\partial}{\partial\eta^{B}_{2}}\frac{\partial}{\partial\eta^{A}_{3}}\frac{\partial}{\partial\eta^{B}_{4}}\mathcal{P}_{\mathcal{N}=2,\mathcal{S}}^{\{H_{i}\}}(x,\eta^{C}_{i})
\\
=&A_{3}^{\psi^{H_{1}-\frac{1}{2}}_{A},\psi^{H_{2}-\frac{1}{2}}_{B},\mathcal{W}_{S+1}}\cdot A_{3}^{\psi^{H_{1}-\frac{1}{2}}_{A},\psi^{H_{2}-\frac{1}{2}}_{B},\mathcal{W}_{S+1}}+A_{3}^{\psi^{H_{1}-\frac{1}{2}}_{A},\psi^{H_{2}-\frac{1}{2}}_{B},\mathcal{W}_{S}}\cdot A_{3}^{\psi^{H_{1}-\frac{1}{2}}_{A},\psi^{H_{2}-\frac{1}{2}}_{B},\mathcal{W}_{S}}
\\
&+A_{3}^{\psi^{H_{1}-\frac{1}{2}}_{A},\psi^{H_{2}-\frac{1}{2}}_{B},\mathcal{W}_{S-1}}\cdot A_{3}^{\psi^{H_{1}-\frac{1}{2}}_{A},\psi^{H_{2}-\frac{1}{2}}_{B},\mathcal{W}_{S-1}}+A_{3}^{\psi^{H_{1}-\frac{1}{2}}_{A},\psi^{H_{2}-\frac{1}{2}}_{B},\phi_{\textbf{P}(CD)}^{(I_{1},\ldots,I_{2S})}  }\cdot A_{3}^{\psi^{H_{3}-\frac{1}{2}}_{A},\psi^{H_{4}-\frac{1}{2}}_{B},\phi^{(CD)}_{\textbf{P}(I_{1},\ldots,I_{2S})}}\, .
\end{split}
\end{equation}

\noindent\textbf{Generalize super polynomial}\\
We have super polynomial in $\mathcal{N}=1$, $\mathcal{N}=2$ SUSY. These results can be easily generalize to extended supersymmetry by replace the Grassmann delta function. For $\mathcal{N}$ -extended supersymmetry, $m\delta^{(4)}(Q^{\dagger}_{4pts})$ is replaced with $m^{\mathcal{N}}\delta^{(2\mathcal{N})}(Q^{\dagger}_{4pts})$ in the super polynomial. The $\mathcal{N}$ super polynomials are
\begin{equation}
\label{eq:4ptn}
\begin{split}
&\boxed{\mathcal{P}^{\{H_{i}\}}_{\mathcal{N},\mathcal{S}}(x,\eta_{i}^{\mathcal{N}})=m^{\mathcal{N}}\delta^{(2\mathcal{N})}(Q^{\dagger}) g(H_{i},\mathcal{S})\mathcal{J}^{H_{1}-H_{2},H_{3}-H_{4}}_{\mathcal{S}}\left(-x \right)
}.
\end{split}
\end{equation}

\section{The supersymmetric EFThedron }\label{sec:EFT}
In this section, we will consider difference R-symmetry SUSY theory in low energy limit. The low energy means our kinematics $s,t$ are much smaller than UV mass scale $\Lambda_{UV}$. We further reduce super amplitude to component amplitude and show how R-symmetry bound EFT couplings. Consider gluon MHV amplitude $\mathcal{A}^{N}$ reduce from super-amplitude:
\eq
\mathcal{A}^{\mathcal{N}}(1^+,2^-,3^+,4^-)=\SB{1 3}^2\AB{2 4}^2\mathcal{A}^{\mathcal{N}}(s,t).
\eqe
We factor out common factor of spinor $\SB{1 3}^2\AB{2 4}^2$ in each $\mathcal{A}^{\mathcal{N}}(1^+,2^-,3^+,4^-)$ amplitudes and focus on improvement of $\mathcal{A}^{\mathcal{N}}(s,t)$ under SUSY increase. The low energy expansion of $\mathcal{A}^{\mathcal{N}}$ is
\eq
\mathcal{A}^{\mathcal{N}}(s,t)|_{s,t\ll \Lambda_{UV}}=(\frac{a}{st}+\frac{b}{s}{+}\frac{b}{t}+\sum_{k,q\geq 0} g_{k,q}s^{k{-}q}t^q)\,.
\eqe
Here we assume there is crossing symmetry $g_{k,q}=g_{k,k-q}$ in amplitude. 
$g_{k,q}$ correspond to the effective couplings under low energy limit which can be extrapolated by contour integral around zero in the complex $s$-plane
\eq
g_{k,q}=\frac{1}{q!}\frac{d^q}{dt^q}\left(\frac{i}{2p}\oint\frac{ds}{s^n} \mathcal{A}(s,t)\right)\bigg|_{t=0}\,.
\eqe
We can deform the integral and catch the poles and discontinuities on real s-axis from 
\eq
\label{jocobiexpand}
\mathcal{A}^{\mathcal{N}}(1^+,2^-,3^+,4^-)|_{s\rightarrow m^2}=\sum_{\ell} c_\ell\frac{\AB{2 4}^{\mathcal{N}} \mathcal{P}_{\ell}^{\alpha,\beta}(\cos\theta)}{s-m^2},\quad \cos\theta=1{+}\frac{2t}{m^2}\, , 
\eqe
 where $\alpha,\beta$ correspond to super-field helicity weight and for MHV amplitude $\mathcal{M}(\bar{\Phi}^{+1},\Phi^{\mathcal{N}/2-1},\bar{\Phi}^{+1},\Phi^{\mathcal{N}/2-1})$ we consider is
 \eq
 \alpha=H_1-H_2=2-\frac{\mathcal{N}}{2},\quad\beta=H_3-H_4=2-\frac{\mathcal{N}}{2}.
 \eqe
We expand (\ref{jocobiexpand}) in Taylor series and factor out common spinor factor. The coefficients of expanding super polynomial are $v_{\ell,q}^{\alpha,\beta}$
\eq
\AB{2 4}^{\mathcal{N}} \mathcal{P}_{\ell}^{\alpha,\beta}(1{+}\frac{2t}{m^2})=\SB{1 3}^2\AB{2 4}^2\sum_{\ell}v_{\ell,q}^{\alpha,\beta}t^{q}.
\eqe
At fix power of $k$, we have
\eq
\vec{g}_k=\sum \left(\begin{array}{c} g_{k,0} \\ g_{k,1} \\ \vdots \\ g_{k,n}\end{array}\right)\in\sum_{a} c_a \vec{\mathcal{P}}_{\ell_a}^{\alpha,\beta},\quad 
 \vec{\mathcal{P}}_{\ell}^{\alpha,\beta}=\left(\begin{array}{c} v_{\ell, 0}^{\alpha,\beta} \\ v_{\ell, 1}^{\alpha,\beta} \\ \vdots \\ v_{\ell, n}^{\alpha,\beta}\end{array}\right)\,,
\eqe
where $a$ labels the spectrum of the UV states and $c_a>0$. This imply vector $\vec{g}_k$ lie inside the convex hull of the super polynomial vector $\vec{\mathcal{P}}_{\ell}^{\alpha,\beta}$. Due to positivity properties of $\vec{\mathcal{P}}_{\ell}^{\alpha,\beta}$, its convex hull is a cyclic polytope. The boundary of cyclic polytope is constructed by adjacent pair of $\vec{\mathcal{P}}_{\ell}^{\alpha,\beta}$, inside the convex hull have
\eq
\langle\vec{g}_k, \vec{\mathcal{P}}_i, \vec{\mathcal{P}}_{i+1}...\vec{\mathcal{P}}_j, \vec{\mathcal{P}}_{j+1}\rangle>0.
\eqe
We use cyclic polytope constrain bound vector $\vec{g}_k$ in each order $k$ and comparing the difference results between $\mathcal{N}=0\sim4$.
\subsection{Expansion}
\begin{figure}[h!]
  \centering
  \begin{subfigure}[b]{0.45\linewidth}
    \includegraphics[width=\linewidth]{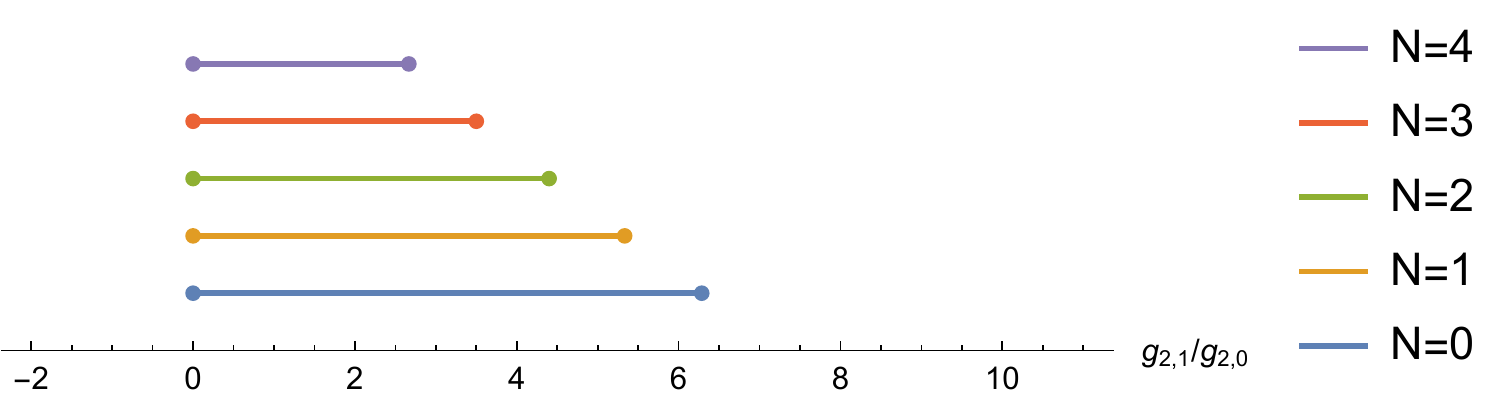}
    \caption{Cyclic-polytope condition on $g_{2,1}/g_{2,0}$.}
    \label{N2}
  \end{subfigure}
  \begin{subfigure}[b]{0.45\linewidth}
    \includegraphics[width=\linewidth]{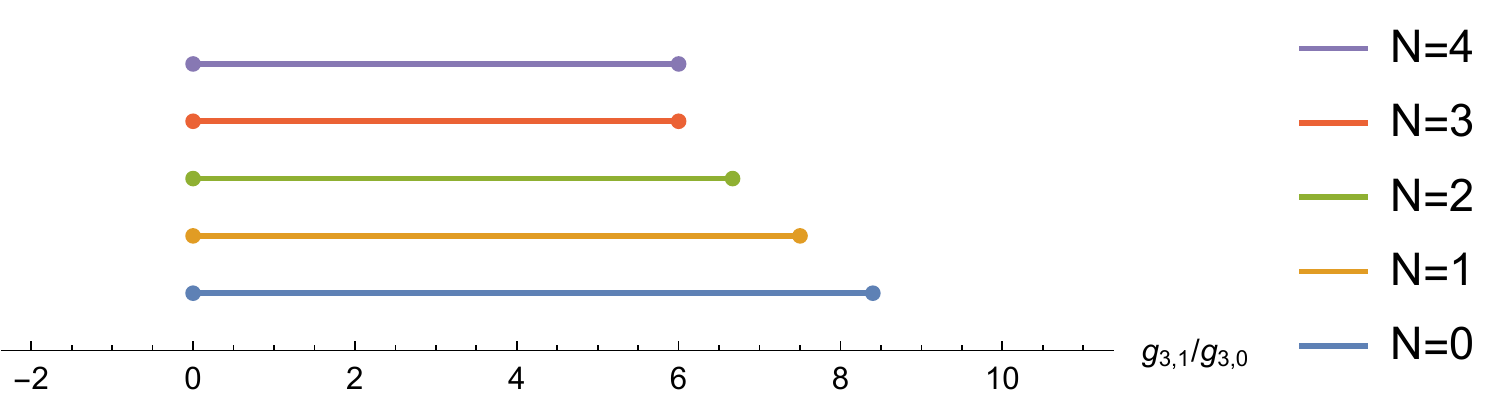}
    \caption{Cyclic-polytope condition on $g_{3,1}/g_{3,0}$.}
    \label{N3}
  \end{subfigure}
  \caption{Cyclic-polytope condition. $\mathcal{N}$ label R-symmetry group}
  \label{N45}
\end{figure}

To expansion $k=2$, we use cyclic-polytope condition constrain $\vec{g}_{2}=(1,g_{2,1}/g_{2,0},g_{2,2}/g_{2,0})$ in projective space. Under the crossing symmetry, $\vec{g}_{2}$ is only one variable vector $\vec{g}_{2}=(1,g_{2,1}/g_{2,0},1)$. We consider super polynomials in each $\mathcal{N}$, and list the most constrained condition from cyclic-polytope for $g_{2,1}/g_{2,0}$:
\eqa
\nonumber \mathcal{N}=0 :&\quad \langle\vec{g}_2,\vec{v}_3,\vec{v}_4\rangle\geq0\quad\rightarrow\quad &0\leq g_{2,1}/g_{2,0} \leq44/7\\
\nonumber \mathcal{N}=1 :&\quad \langle\vec{g}_2,\vec{v}_2,\vec{v}_3\rangle\geq0\quad\rightarrow\quad&0\leq g_{2,1}/g_{2,0} \leq16/3\\
\nonumber \mathcal{N}=2 :&\quad \langle\vec{g}_2,\vec{v}_2,\vec{v}_3\rangle\geq0\quad\rightarrow\quad&0\leq g_{2,1}/g_{2,0} \leq22/5\\
\nonumber \mathcal{N}=3 :&\quad \langle\vec{g}_2,\vec{v}_1,\vec{v}_2\rangle\geq0\quad\rightarrow\quad&0\leq g_{2,1}/g_{2,0} \leq7/2\\
 \mathcal{N}=4 :&\quad \langle\vec{g}_2,\vec{v}_1,\vec{v}_2\rangle\geq0\quad\rightarrow\quad&0\leq g_{2,1}/g_{2,0} \leq8/3.
\eqae
We find that with higher order $\mathcal{N}$, $g_{2,1}/g_{2,0}$ is bounded in tighter region. The result also show in fig.\ref{N2}.

The same procedure can go to $k=3$ expansion. In $k=3$, $\vec{g}_3=(1,g_{3,1}/g_{3,0},g_{3,2}/g_{3,0},g_{3,3}/g_{3,0})$ still be an one variable vector $(1,g_{3,1}/g_{3,0},g_{3,1}/g_{3,0},1)$. One can read out cyclic-polytope constrain and have upper bound for $\{\mathcal{N}{=}0,\mathcal{N}{=}1,\mathcal{N}{=}2,\mathcal{N}{=}3,\mathcal{N}{=}4\}=\{42/5,15/2,20/3,6,6\}$ as show in fig.\ref{N3}. The region become smaller with increasing $\mathcal{N}$. In $k=3$, $\mathcal{N}=3$ and $\mathcal{N}=4$ share same boundary. It is not necessary for them have same boundary, one can find difference boundaries in others $k$.  

In $k=4$ and $k=5$, there are both two independent variables. We show the constrain in fig.\ref{N45}.
\begin{figure}[h!]
  \centering
  \begin{subfigure}[b]{0.3\linewidth}
    \includegraphics[width=\linewidth]{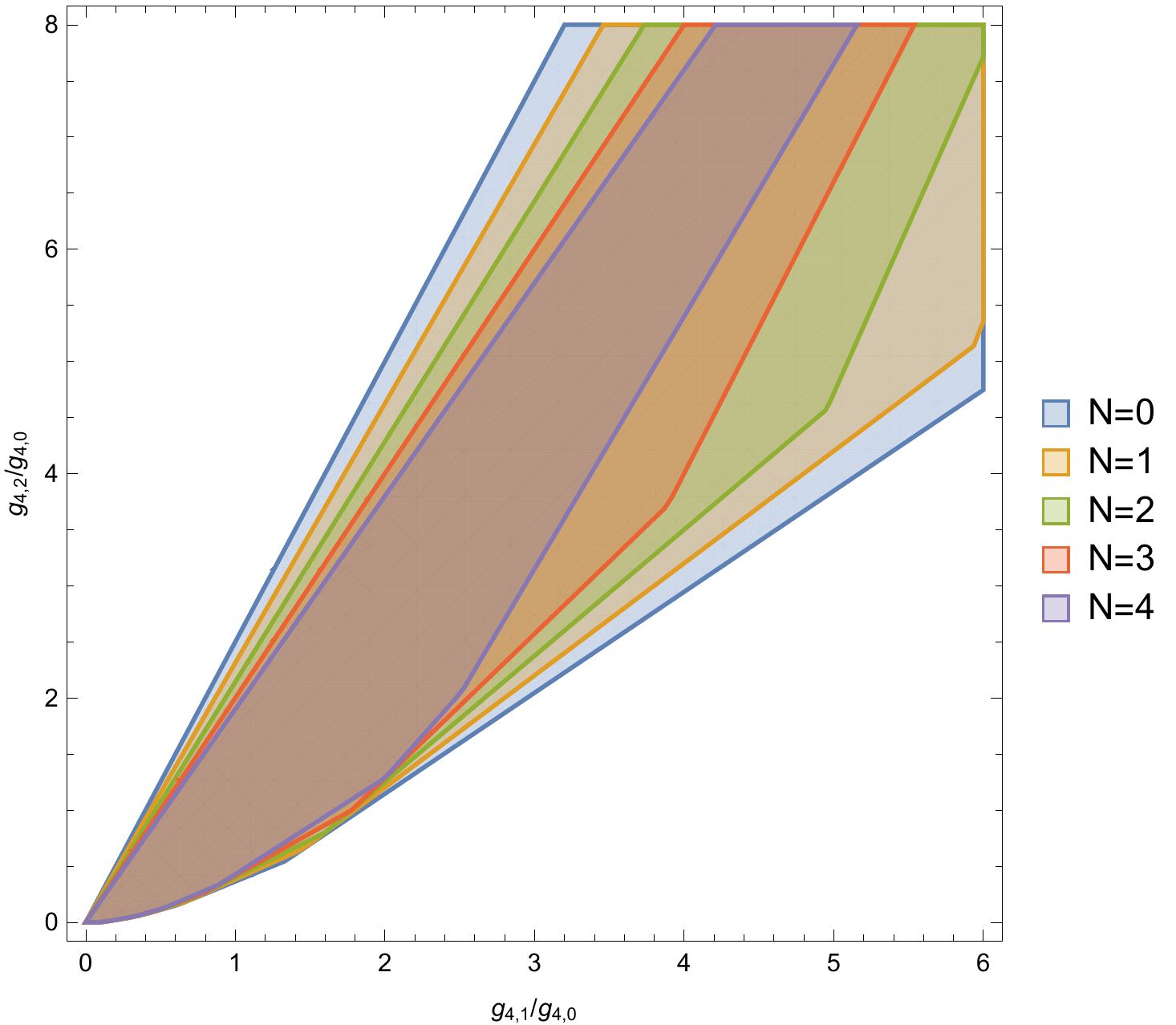}
    \caption{Cyclic-polytope condition on $(g_{4,1}/g_{4,0},g_{4,2}/g_{4,0})$}
  \end{subfigure}
  \begin{subfigure}[b]{0.3\linewidth}
    \includegraphics[width=\linewidth]{N5.pdf}
    \caption{Cyclic-polytope condition on $(g_{5,1}/g_{5,0},g_{5,2}/g_{5,0})$}
  \end{subfigure}
  \caption{Cyclic-polytope condition. $\mathcal{N}$ label R-symmetry group.}
  \label{N45}
\end{figure}
The region become smaller is reflected the fact, the decomposition of $\mathcal{N}$ super polynomial is positive sum of $\mathcal{N}{-}1$ polynomials. It can be easily show from recurrence relation 
\begin{equation}
\begin{split}
\frac{\partial}{\partial\eta^{A}_{2}}\frac{\partial}{\partial\eta^{A}_{4}}\mathcal{P}^{\mathcal{N}}_{\{H_{i}\},S}\vert_{\eta^{A}=0}=m^{2}\mathcal{P}^{\mathcal{N}-1}_{\{h_{i}\},S+\frac{1}{2}}+\frac{m^{2}(S-H_{3}+H_{4})(S-H_{1}+H_{2})}{\left( 2S\right) \left( 2S+1\right)  }\mathcal{P}^{\mathcal{N}-1}_{\{h_{i}\},S-\frac{1}{2}}\, .
\end{split}
\end{equation}
RHS have positive coefficients when we identify $H_1=H_3, H_2=H_4$ which means $\mathcal{N}$ super polynomial is living inside of $\mathcal{N}{-}1$ convex hull. As we know the boundary of $\mathcal{N}$ convex hull are living inside of $\mathcal{N}{-}1$, the $\mathcal{N}$ space must smaller than $\mathcal{N}{-}1$.

\section{Conclusion}
In this paper, we proposed an super Poincare invariant expansion basis for the residue and discontinuity of four-point amplitudes in four dimensions. These super spinning polynomials are given in terms of algebraic Jacobi functions, whose recurrence relations reflect the decomposition of the SUSY multiplet in terms of components. These polynomials were identified through gluing of three-point super amplitudes constructed out of massive and massless on-shell super space. 

As an application, we use dispersion relations to derive bounds on EFT coefficients. By expanding the imaginary part of the amplitude on these super spinning polynomials, we derive bounds that reflect the underlying supersymmetric UV completion.  
\section{Acknowledgements}
We would like to thank Yu-tin Huang for enlightening discussions. J-Y Liu  and  Z-M You are supported by MoST Grant No. 106-2628-M-002-012-MY3.

\label{sec:Conclusion}

\appendix
\section{Conventions}
\label{sec:spinor-helicity}
The convention we used in the central of mass frame.  Spinor brackets are related to center of mass $m$ and scattering angle $\theta$:
 \begin{equation}
 \label{eq:k}
 \begin{split}
&\vert 1\rangle = m^{\frac{1}{2}} 
\left( \begin{array}{cc} 
    1&\\ 
    0&  
\end{array}\right) 
\,,\quad\vert 2\rangle =  m^{\frac{1}{2}} 
\left(\begin{array}{cc} 
    0 &\\ 
    1 &  
\end{array}\right) 
\\
&\vert 3\rangle = i m^{\frac{1}{2}} 
\left(\begin{array}{cc} 
    sin\frac{\theta}{2}&\\ 
     -cos\frac{\theta}{2}&  
\end{array}\right) 
\,,\quad\vert 4\rangle = i m^{\frac{1}{2}} 
\left(\begin{array}{cc} 
     cos\frac{\theta}{2} &\\ 
    sin\frac{\theta}{2} &  
\end{array}\right) .
\end{split}
\end{equation}
Levi-Civita tensor is
\eq
\epsilon^{a\dot{a}}=
\left( \begin{array}{cc} 
    0&1\\ 
    -1&0  
\end{array}\right) .
\eqe
Contraction of spinor brackets:
\eqa
\nonumber&\langle12 \rangle= m,\quad  &\langle34\rangle= - m\\
\nonumber&\langle13\rangle=  -im cos(\frac{\theta}{2}),\quad&\langle24\rangle =  -im cos(\frac{\theta}{2})\\
\label{eq: angle}&\langle14\rangle= im  sin(\frac{\theta}{2}),\quad&\langle23\rangle =   -im sin(\frac{\theta}{2}) .
\eqae
The Mandelsteine variables define as
\begin{align}
\nonumber s=-(P_{1}+P_{2})^2,\quad t=-(P_{1}+P_{4})^2,\quad u=-(P_{1}+P_{3}	)^2\\
\Rightarrow s=m^2,\quad t=-m^{2}\frac{(1-cos\theta)}{2},\quad u=-m^2\frac{(1+cos\theta)}{2} .
\end{align}


\end{document}